\documentclass[nocopyrightspace]{sigplanconf}

\usepackage{amsmath}

\usepackage{amssymb}

\usepackage{graphicx}

\usepackage{subfig}

\usepackage{float}

\usepackage{caption}

\usepackage{framed}

\usepackage{multirow}

\usepackage{balance}

\usepackage{color}

\usepackage{hyperref}

\usepackage[all]{hypcap}

\newenvironment{compact_enum}
{\setlength{\leftmargini}{1em}
\begin{enumerate}
  \setlength{\labelsep}{.3em} 
  \setlength{\itemsep}{.4em}
  \setlength{\parskip}{0pt}
  \setlength{\parsep}{0pt}}
{\end{enumerate}}

\newenvironment{compact_item}
{\setlength{\leftmargini}{1em}
\begin{itemize}
  \setlength{\labelsep}{.3em} 
  \setlength{\itemsep}{.4em}
  \setlength{\parskip}{0pt}
  \setlength{\parsep}{0pt}}
{\end{itemize}}

\def\semijoin{\mbox{$\mathrel{\raise1pt\hbox{\vrule height5pt depth0pt\hskip-1.5pt$>$\hskip -2.5pt$<$}}$}}

\begin{document}

\conferenceinfo{DBPL 2013}{August 30, 2013, Riva del Garda, Trento, Italy.} 
\copyrightyear{2013} 
\copyrightdata{[to be supplied]}

\title{Declarative Ajax Web Applications through \\ SQL++ on a Unified Application State%
\titlenote{Supported by NSF III-1018961 and NSF III-1219263, PI'd by Prof Papakonstantinou who is a shareholder of App2you Inc, which commercializes outcomes of this research.}
}

\authorinfo{Yupeng Fu}
           {UCSD}
           {y4fu@cs.ucsd.edu}
\authorinfo{Kian Win Ong}
           {UCSD}
           {kianwin@cs.ucsd.edu}
\authorinfo{Yannis Papakonstantinou}
           {UCSD}
           {yannis@cs.ucsd.edu}

\maketitle

\begin{abstract}
Implementing even a conceptually simple web application requires an inordinate amount of time. FORWARD addresses three problems that reduce developer productivity: (a) {\em Impedance mismatch} across the multiple languages used at different tiers of the application architecture. (b) {\em Distributed data access} across the multiple data sources of the application (user input of the browser page, session data in the application server, SQL database etc). (c) {\em Asynchronous, incremental modification of the pages}, as performed by Ajax actions.

FORWARD belongs to a novel family of web application frameworks that attack impedance mismatch by offering a single unifying language. FORWARD's language is {\em SQL++}, which is SQL with necessary extensions for semi-structured data. FORWARD's architecture is based on two novel cornerstones: (a) A {\em Unified Application State} (UAS), which is a virtual database over the multiple data sources. The UAS is accessed via distributed SQL++ queries, therefore resolving the distributed data access problem. (b) Declarative page specifications, which treat the data displayed by pages as {\em rendered SQL++ page queries}. The resulting pages are automatically incrementally modified by FORWARD. User input on the page becomes part of the UAS. 

We show that SQL++ is suited for the semi-structured nature of web pages and captures the key data model aspects of two important data sources of the UAS: SQL databases and JavaScript components. We show that simple markup is sufficient for creating Ajax displays and for modeling user input on the page as UAS data sources. Finally, we discuss the page specification syntax and semantics that are needed in order to avoid race conditions and conflicts between the user input and the automated Ajax page modifications.

FORWARD has been used in the development of eight commercial and academic applications. An alpha-release web-based IDE (itself built in FORWARD) enables development in the cloud.

\end{abstract}

\category{H.3.3}{Programming Languages}{Language Constructs and Features - Frameworks}
\category{H.2.8}{Information Systems}{Database Management - Applications}

\terms
Languages

\keywords
Declarative, AJAX, Web Application, Web Application Framework, SQL, FORWARD

\section{Introduction}
\label{sec:introduction}

Implementing web applications requires an inordinate amount of time even for an experienced developer.
This is because web applications have a three-tier application architecture, each tier potentially running on a separate machine: (a) the visual tier in the browser, (b) the application logic tier in the application server, and (c) the data tier in the database. Multiple mundane low-level tasks need to be performed in a coordinated fashion across all three tiers. The resulting problems (as follows) reduce developer productivity:

\begin{compact_enum}
\item {\em Impedance mismatch} arises because each tier uses different languages (and data models) \cite{links-fmco-06,hilda-icde-06}. The visual tier uses DOM (XHTML) / JavaScript (JSON); the application logic tier uses an application programming language, such as Java, Python or PHP (objects); the data tier uses SQL (tables). To display a page, mundane and error-prone code is needed to translate SQL tables into Java objects, and then into HTML / JavaScript objects. To implement an action invocation, more code is needed to translate in the opposite direction. 

\item The developer engages in ad-hoc {\em distributed data access} across the multiple data sources and machines that the application state is distributed over \cite{status-quo-cidr-13,hilda-www-07}. For example, suppose a user selects a location in a browser-side street map, and an action needs to compare if the selection is equal to some location stored in the database. The developer has to programatically retrieve the location from the browser-side component, marshall it across the network, unmarshall it on the server, and parameterize a query to send to the database. This ad-hoc distributed data access is primarily due to the application state being distributed across multiple machines: such complexities still occur even if the database has the same data model as the browser (as is the case for JSON databases such as MongoDB).

\item In order to improve latency and user experience when refreshing complex pages, the developer handcodes Ajax optimizations that imperatively perform {\em asynchronous, incremental modification} from the old page to the new page \cite{forward-sigmod-10}. Instead of re-computing the page from scratch, an action issues a ``delta" SQL query retrieving the subset of data needed for refresh, performs incremental computation on the server, sends diffs to the browser, and finally uses the incremental rendering methods of the DOM and JavaScript components. Such event-driven programming to perform incremental computations is well-known to be error-prone and laborious, as it requires the developer to correctly assess the data flow dependencies on the page, and implement for each action on the page additional custom code that correctly transitions the application from one consistent state to another. This is further compounded by asynchronicity: since incremental modifications of the page execute in non-deterministic order, the developer has to reason about how each action interacts with other potentially concurrent actions.

\end{compact_enum}

To mitigate the productivity sink from these mundane low-level tasks, practitioners have relied primarily on web application frameworks, such as Ruby-on-Rails, Django and Microsoft ASP.NET. Each framework is built on top of a mainstream programming language of the application logic tier (Wikipedia tabulates 110 web application frameworks across 10 mainstream programming languages \cite{web-application-frameworks-wikipedia-13}), and focuses on providing libraries that bridge pair-wise gaps between the application logic tier and the other two tiers. For example, an Object-Relational Mapper (ORM) library mitigates the impedance mismatch between the programming language and SQL database, a serialization library simplifies sending JSON across the network, a JavaScript utility library (e.g. JQuery) facilitates navigating and updating the DOM for incremental refresh, etc.  
Since libraries focus on mitigating point issues at a particular system boundary instead of holistically handling issues across multiple tiers and data sources, it is up to the developer to coordinate these separate libraries. Furthermore, a developer cannot rely on consistent semantics across libraries when verifying whether edge cases are correctly handled end-to-end over the three tiers. Ultimately, a developer still needs to be aware of the differences between the underlying languages and data models when building Ajax web applications.

Recognizing that libraries are not sufficient to provide conceptually clean and simple abstractions for Ajax web application programming, the database and programming language communities have proposed novel frameworks such as Links \cite{links-fmco-06} and Hilda \cite{hilda-icde-06}, each based on a single language that spans all three tiers. In the same spirit, we present FORWARD \cite{forward-sigmod-10,forward-cidr-11}, an Ajax web application framework that provides a holistic abstraction through a single language SQL++, which is SQL with necessary semi-structured extensions to support the nesting and heterogeneity that pages typically exhibit. FORWARD further improves upon Links and Hilda with two novel cornerstones in its architecture: the {\em unified application state} (UAS) encapsulating multiple data sources, and declarative pages via {\em rendered queries}. A FORWARD application comprises:

\begin{compact_enum}
\item a {\em Unified Application State (UAS) specification}%
, which specifies a virtual database comprising nested heterogeneous tables that integrate data from multiple data sources of varying data models, including SQL databases, sessions, URL parameters, the HTML DOM and JavaScript components.
\item {\em page specifications} (e.g. Figures~\ref{fig:page-static}, \ref{fig:page-report}, \ref{fig:page-event}), which declaratively specify the data of a page using SQL++ queries that execute over the UAS. The rendering of data is also specified declaratively using template markup, which goes beyond HTML to support JavaScript components with rich behavior and functionality, such as maps, calendars and bar charts. Since page specifications are declarative, the incremental modification of pages is handled automatically by FORWARD's optimizations.
\item {\em action specifications} (e.g. Figure~\ref{fig:action}), which are implemented in PL/SQL++, the analogously extended version of the procedural language of SQL databases. Thus, application state is manipulated using basic control flow constructs (such as conditionals, loops and functions) and  {\tt INSERT}, {\tt UPDATE} and {\tt DELETE} statements on the UAS.
\end{compact_enum}

The FORWARD compiler performs static analysis on the above specifications and performs automatic optimizations.

Since SQL++ offers a single language and data model, there is no impedance mismatch. Instead of ad-hoc distributed data access, the UAS allows the developer to write SQL++ queries and updates in a {\em location transparent} fashion: FORWARD's distributed query processor will compile a SQL++ query into separate, efficient queries on respective data sources \cite{forward-data-access-ucsd-13}. As a result of declarative pages, the contents of a page are specified without regard for how they will be asynchronously, incrementally modified. When changes occur in the underlying data dependencies of a page, FORWARD's optimizations will automatically provide incremental modification of pages through incremental view maintenance and incremental rendering \cite{forward-sigmod-10}. Thus, FORWARD simultaneously provides ease of use for the developer, as well as efficient performance through automatic optimizations.

In designing the single language of the framework to meet the needs of the UAS and page-as-rendered-query cornerstones, we choose to base it on SQL because of SQL's:
\begin{compact_item}
\item {\bf Familiarity to developers: } Since the primary persistent data stores of web applications are SQL databases, web developers are already familiar with using SQL to query and update the application state.

\item {\bf Declarativeness: } While rendered queries provide page semantics that are easy to understand, FORWARD also evaluates pages efficiently so that performance is comparable to (or better than) incremental computations a developer would otherwise provide manually. As a declarative language, SQL provides opportunities for automatic performance optimizations during page evaluation, including: incremental evaluation of queries using incremental view maintenance \cite{forward-sigmod-10}, rewriting tuple-at-a-time queries into more efficient normalized-sets queries \cite{forward-data-access-ucsd-13}, and bounding communication costs for distributed query processing \cite{forward-data-access-ucsd-13}.

\item {\bf Expressiveness: } We observe that the data of pages are typically (a) associations between different data sources, which are handled by joins in a distributed query (b) analytics and aggregations over source data, which are handled by {\tt GROUP BY} and aggregation functions. Moreover, actions typically comprise create, read, update and delete (CRUD) data operations, which are handled by {\tt INSERT}, {\tt UPDATE} and {\tt DELETE} statements. SQL's expressiveness makes the common cases easy, and provides extensibility points for more complex operations via user-defined functions (UDFs).

As a data point from our system implementation, the most complex FORWARD application deployed in production has 18 pages, 207 actions and 80 UDFs. 72 of the UDFs are for modularization/reusability purposes and implemented in a few lines of SQL++; the remaining 8 required Java. The fact that Java UDFs comprise only 3\% of all actions/UDFs suggests that SQL++ and PL/SQL++ have very good fit with common cases. We comment in the Future Work section on the reasons for Java UDFs and how we can further increase the coverage provided by SQL++.

\item {\bf Inter-operability with data model of JavaScript components: } In the visual tier, most JavaScript components exchange data using JSON values. For the unifying data model of the UAS, we observe that the data model of SQL is close to JSON, since both make the distinction between collections (tables versus arrays) and tuples (rows versus object literals, i.e. sets of name-value pairs). In contrast, XML does not make such a distinction since it only recognizes collections of elements, which makes it less suitable for the unifying data model.

\item {\bf Maturity: } Utilizing SQL allows us to leverage mature research techniques, by formulating reductions from FORWARD optimization and static analysis problems to extensions/specializations of respective problems in the data management community, including schema mappings, updateable views, distributed query processing and incremental view maintenance.

\end{compact_item}

In this paper, we present a system overview of FORWARD, illustrate its language by example, and discuss how the characteristics of modern Ajax web applications are captured. In particular, we highlight the following novel contributions to language design that are improvements/extensions over early prototypes \cite{forward-sigmod-10,forward-cidr-11}:

\begin{compact_enum}

\item {\bf SQL++ data model that captures state of JSON components: } To bridge the gap between the data models of two important data sources of the UAS, namely tables of SQL databases and JSON of JavaScript components, we design a unifying data model for SQL++. We describe how a SQL++ value is isomorphic to a JSON value, thereby showing that one may think of the FORWARD data model as either SQL++ or JSON. In particular, we extend SQL tables to capture the nesting, ordering and heterogeneity specific to JSON. Vice versa, we consider an extended JSON that captures richer scalar types. Separately, values in the HTML DOM are handled as special cases. (Sections~\ref{sec:data-model}, \ref{sec:query})

\item {\bf Declarative template language for HTML and JavaScript components of pages: } Pages are declaratively specified as rendered queries, such that data are specified with SQL++ queries and renderings are specified with HTML/JSON template markup. In particular, JavaScript components can be easily specified through JSON template markup, even when the components provide only programmatic interfaces natively. This is enabled by modelling a page instance with a SQL++ value. FORWARD automatically propagates changes from the page instance to JavaScript components and vice versa, and automatically optimizes by using efficient incremental rendering methods of the JavaScript components. (Section~\ref{sec:page-static})

\item {\bf Modelling user input versus visualizations for asynchronous actions: } A page can be conceptually thought of as a join of two distinct parts: a stateful part for user input, and a stateless part for visualizations that are displayed to the user but not modified directly via user activity. To guarantee that asynchronous actions will not cause page refreshes that conflict with user input, FORWARD models the page such that in a page instance each value is either (a) a {\em mutable value} stored in the UAS or (b) an {\em immutable value} that is the result of a rendered query over the UAS. Whereas mutable values may be concurrently written by the user and multiple asynchronous actions, immutable values are always consistent with respect to an up-to-date UAS. (Sections~\ref{sec:page-event}, \ref{sec:action})

\item {\bf System implementation: } To validate our language design, we have implemented FORWARD and used it to build multiple applications, including commercial ones. A prominent FORWARD application is an alpha-release IDE that allows developers to create other hosted FORWARD applications in the cloud, made available at \url{http://forward.ucsd.edu/}. Furthermore, FORWARD is currently deployed in commercial applications, of which a characteristic one is BioHeatMap (\url{http://www.bioheatmap.com}), an analytics application currently utilized by two pharmaceutical companies.

\end{compact_enum}

\section{Syntax and Semantics}
\label{sec:semantics}

\begin{figure}[htbp]
\begin{center}
  \fbox{\includegraphics[width=3in]{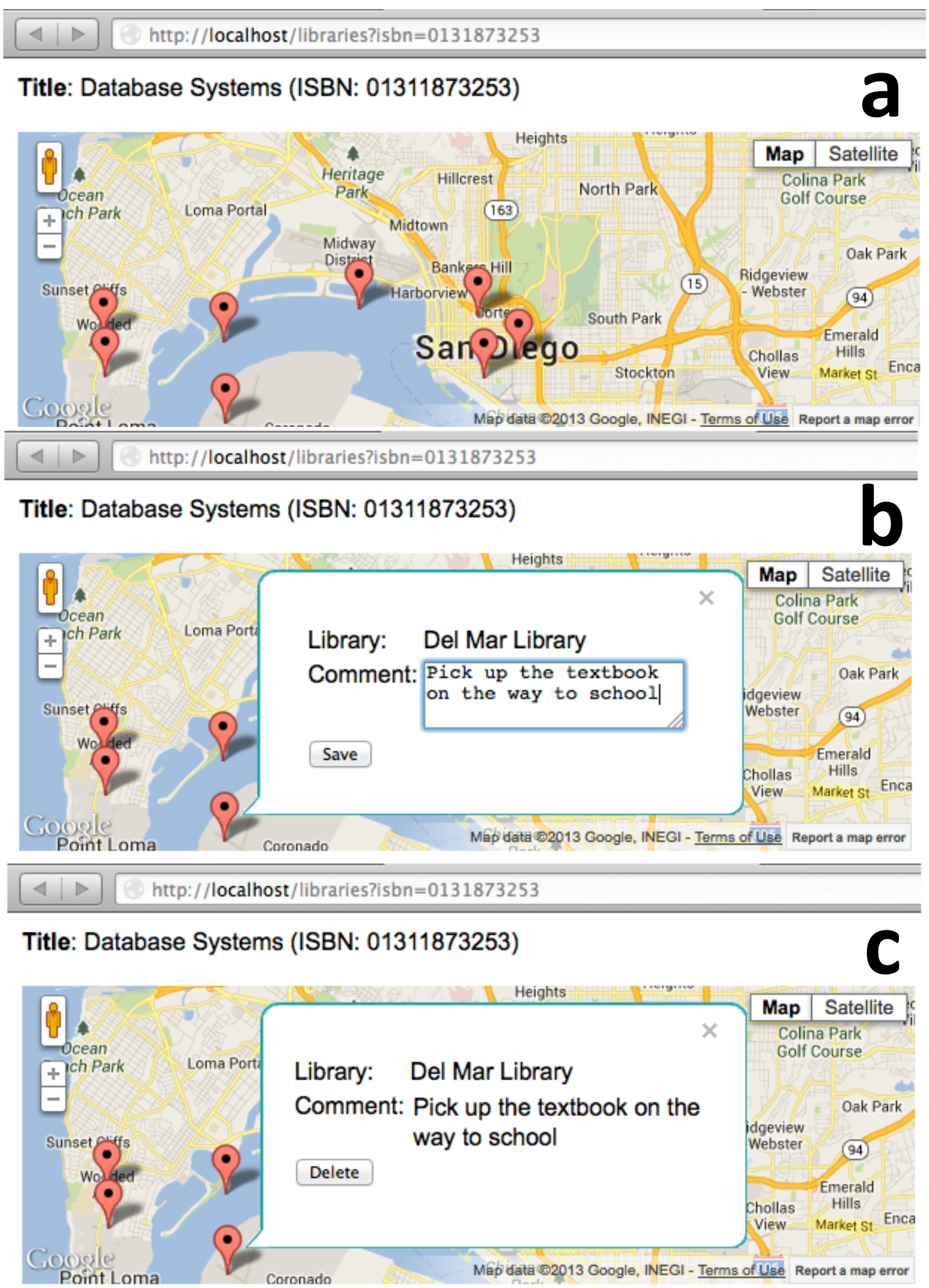}}
\caption{Map of libraries with copies of a book}
\label{fig:screenshot}
\end{center}
\end{figure}

The syntax and semantics are illustrated by using FORWARD to build a simple data-driven Ajax web application. Figure~\ref{fig:screenshot} shows the sample application which displays the availability of books at libraries: (a) Given a book's ISBN number as a URL parameter, the libraries with available copies are visualized as markers on a street map. (b) The user can select a marker, comment on the particular location of the book, and save the note in the session. (c) A saved note is displayed, and can be subsequently deleted. Notice that the URL stays the same across user activity, as incremental page modifications are performed through Ajax.

Throughout the paper, code examples to be highlighted are provided inline, whereas the Appendix provides the complete 117 lines implementing the running example. 

In FORWARD, an application comprises specifications for the UAS, pages and actions. When the FORWARD runtime interpreter executes the application, an {\em action-page cycle} occurs in response to user interaction, such that each cycle comprises an action invocation that both accesses and mutates the UAS, and a page evaluation that only accesses the UAS without mutating it. Consider when a user enters a URL in a blank browser window, such as \url{http://localhost/libraries?isbn=0131873253} to initiate Figure~\ref{fig:screenshot}a. The runtime finds the {\tt libraries} action (Appendix Figure~\ref{fig:appendix-libraries-action}) associated with the URL and invokes it. The action reads and writes the UAS, which comprises multiple data sources including SQL databases, the session, page input data, and URL parameters, and possibly invokes functions with side-effects such as sending emails. An action also selects the page specification to be evaluated and displayed next, in this case the {\tt map} page. The runtime then evaluates the selected {\tt map} page specification (Appendix Figure~\ref{fig:appendix-map-page}) to produce a {\em page instance}, during which it enforces that no writes occurs on the UAS, and synchronizes the page instance with the browser.

As the user interacts with the browser, such as zooming the map, selecting a marker or typing in a text area as in Figure~\ref{fig:screenshot}b, the runtime synchronizes user input back into the page instance. The action-page cycle continues when user activity (1) causes navigation to a different URL, in which case the browser operates as if the user has manually entered the URL in the address bar, or (2) fires an event that has been bound to an Ajax action on the page, in which case the address bar's URL remains the same, and the runtime invokes the action asynchronously via XHR (i.e. the XmlHttpRequest API). User input within the page instance become part of the UAS, which enables the action to access them. In Figure~\ref{fig:screenshot}b for example, when the user clicks the {\tt Save} button, the runtime asynchronously invokes Ajax action {\tt save\_note} (Appendix Figure~\ref{fig:appendix-save-note-action}), which reads the user input of the text area and writes it via the UAS into the session. Since Ajax actions are asynchronous by default, multiple Ajax actions on a page can execute concurrently and cause refreshes of the browser in non-deterministic order. For example, after {\tt Save} is clicked, the runtime typically refreshes the browser to Figure~\ref{fig:screenshot}c, but if network delays occur or the action had incurred long-running computations, the user can continue to pan the map or select other markers without stalling.

\paragraph{Comparison with mainstream MVC web frameworks} FORWARD's programming model follows the established Model-View-Controller (MVC) \cite{mvc-joop-88} architecture pattern, which is also followed by Ruby-on-Rails, Django, ASP.NET and other mainstream web frameworks. A MVC web application is modularised into models (analogous to the UAS), UI views (pages) and controllers (actions). For Ajax support, however, a controller is typically burdened with incrementally modifying the displayed UI view in order to reflect the model changes incurred by the action. 

In FORWARD, a key architecture distinction from prior MVC web frameworks is that an action is not responsible for the modification of an Ajax page. Rather, the framework automatically reflects on the page the changes made in the UAS. For example, clicking the {\tt Save} button in Figure~\ref{fig:screenshot}b invokes the {\tt /save\_note} action (Appendix Figure~\ref{fig:appendix-save-note-action}) that saves the note in the session, but does not specify how to remove the text area and add a {\tt Delete} button on the page. As a result, FORWARD actions are typically very concise (Appendix Figures~\ref{fig:appendix-libraries-action}, \ref{fig:appendix-save-note-action} and \ref{fig:appendix-delete-note-action}), making it easy to overview the changes they effect on the application state.

In the following sections, we illustrate syntax and semantics by describing how to specify the UAS, pages and actions of the sample application. Section~\ref{sec:uas} describes the UAS specification and SQL++, Section~\ref{sec:page} describes the page specifications, and Section~\ref{sec:action} describes the action specifications.

\subsection{UAS Specification and SQL++}
\label{sec:uas}

The {\em unified application state (UAS)} offers a uniform SQL++ query interface to both persistent and transient data sources, which include SQL databases, sessions, URL parameters, the HTML DOM and JavaScript components. Conceptually, the UAS is a virtual database of SQL++ values and functions from different data sources. By using SQL++ values as a unifying data model that subsumes both SQL tables and JSON, the UAS resolves impedance mismatch between the SQL database and the page. Furthermore, the UAS enables the developer to write location-transparent SQL++ queries.

Section~\ref{sec:source} presents the {\em source specifications} that a developer provides to map respective sources into the UAS. Subsequently, Section~\ref{sec:data-model} discusses how the SQL++ data model subsumes SQL tables and JSON, and Section~\ref{sec:query} discusses the query language extensions in SQL++.

\subsubsection{Source Specifications}
\label{sec:source}

FORWARD supports different types of data sources through {\em source wrappers}, each of which implements how a SQL++ query (or a limited subset thereof) can be executed in a type of source. Each wrapper optionally uses the metadata of sources to automatically map existing schemas and functions into the UAS. The two types of sources currently supported are SQL databases ({\tt sql}) and in-memory SQL++ values ({\tt memory}), whereas other types that may be supported in the future include spreadsheets, JSON databases (such as CouchDB and MongoDB) and large-scale data services (such as Amazon Redshift and Google BigQuery).

A source specification has a source type, a source name that is used in queries to refer to data in the source, and a options particular to the source type. For a {\tt sql} source, this includes authentication credentials and the database name. FORWARD introspects the database's system catalog to discover tables and stored procedures when it connects to the database during compilation and runtime of the application.

In addition to the source specifications provided by the developer, a FORWARD application is also automatically configured with sources corresponding to data stored in the application server: the {\tt session} source corresponds to the HTTP session, the {\tt url} source (read-only) corresponds to the URL parameters in the browser address bar, and the {\tt http\_headers} source (read-only) corresponds to the headers of a HTTP request. These sources are of type {\tt memory}, i.e. data are represented as SQL++ values that are stored in the memory space of the application server. The {\em lifetime scope} of these transient sources is as determined by application servers and browsers. For example, the {\tt session} source lives for the duration of a HTTP session: All actions and pages invoked within a browser session have access to the same instance of the {\tt session} source, while actions and pages of other browser sessions have their own session instances. The {\tt url} source has the same lifetime scope as JavaScript variables in a browser window: A browser will preserve JavaScript variables across incremental refreshes up until the URL in the browser's address bar changes, upon which the browser will reset the DOM and JavaScript variables and load from scratch the HTML and scripts of the new URL\footnote{Special case: If only the fragment identifier (i.e. after the {\tt \#}) has changed in the URL, a browser will preserve JavaScript variables.}. The {\tt http\_headers} source lives for the duration of an action invocation.

Furthermore, FORWARD automatically provides the {\tt request} source that maps into the user input of a page. This source is inferred from the page specification, and described in Section~\ref{sec:page-event} as part of data binding.

\subsubsection{Data model extensions to subsume SQL tables and JSON}
\label{sec:data-model}

The data model of SQL++ subsumes the data models of two important data sources of the UAS: SQL databases and JavaScript components. Whereas SQL uses tables (i.e. bags of tuples), JavaScript components represent data with JSON, a lightweight data-interchange format based on the subset of JavaScript comprising (a) arrays (b) object literals, i.e. sets of name-value pairs (c) basic scalars i.e. strings, numbers and booleans. Although SQL's data model is distinct from JSON, they share important similarities: SQL tables are similar to JSON arrays, SQL tuples are similar to JSON object literals, and they both have scalars. Since we want an extension of SQL to be the query language and an extension of PL/SQL to be the programming language, we base SQL++'s data model on SQL's since query processing is well-understood on SQL tables, and then extend the data model to subsume both SQL and JSON.

\begin{compact_enum}
\item {\bf Typing:} A SQL {\tt CREATE TABLE} statement simultaneously creates the schema (type) and a value conforming to the schema, whereas JSON does not natively support schemas since JavaScript is dynamically typed. To support both SQL and JSON, a SQL++ value is independent of its schema. Static type-checking is optional, and query processing does not require schemas\footnote{Nevertheless query processing takes advantage of statically known types when they are available.}.
\item {\bf Root:} A SQL query always returns a SQL table, whereas the root of a JSON structure is an array, object literal or scalar. Thus, a SQL++ value is a table, tuple or scalar.
\item {\bf Nesting:} A SQL tuple contains only scalars, whereas a JSON object literal contains arrays, object literals and scalars. Thus, a SQL++ tuple contains any SQL++ value. In particular, this extension enables tables to be recursively nested.
\item {\bf Ordering:} A SQL table is a bag of tuples: its order is considered intentional when it is output by a query with {\tt ORDER BY}, but coincidental/non-deterministic otherwise (also, when it is stored on disk). In contrast, a JSON array is always ordered. In SQL++, a table maintains its order and has metadata indicating whether the order is intentional or coincidental. Each tuple also has its ordinal position as metadata.
\item {\bf Heterogeneity:} A SQL table contains only homogeneous tuples, whereas a JSON array can contain arrays, object literals and scalars, each of which is potentially different from the others. Thus, a SQL++ table is a list of potentially heterogeneous SQL++ tuples. As a special case, an element of a JSON array that is either an array/scalar is represented by a wrapper SQL++ tuple that has a table/scalar as its single value. This single value has the sentinel name {\tt \#}.
\item {\bf Scalars} A JSON scalar is a string, number or boolean, whereas SQL supports additional scalars such as binary data, dates and timestamps. Thus, SQL++ allows extensions for additional scalars.
\end{compact_enum}

\subsubsection{Query language extensions over SQL}
\label{sec:query}

The SQL++ query language is backwards-compatible with SQL, with extensions analogous to those in its data model:

\begin{compact_enum}

\item {\bf Root} Unlike SQL where queries always start with {\tt SELECT} and always output tables, SQL++ queries can be simple expressions. For example, {\tt 1 + 1} outputs a SQL++ scalar.

\item {\bf Nesting} Unlike SQL which restricts the {\tt SELECT} clause to only allow sub-queries producing scalars, in SQL++ the {\tt SELECT} clause can contain any SQL++ sub-query. In particular, this extension enables creating nested tables.

\item {\bf Heterogeneity} Unlike SQL which only provides {\tt UNION} which requires both arguments to be homogeneous (i.e. identical schema), SQL++ also provides {\tt OUTER UNION} which allows both arguments to be heterogeneous (i.e. different schemas). 
\end{compact_enum}

Using SQL++ to produce the data of a page will be further discussed in Section~\ref{sec:page-report}.

\subsection{Page Specifications}
\label{sec:page}

A {\em page specification} declaratively specifies the page that is presented to the user, including displayed visualizations of data (e.g. HTML tables, bar charts), user input that is stored (e.g. HTML forms, zoom level of a map), and how the user can interact with the page (e.g. clicking a button, dragging a map marker). Since the page specification is declarative, the developer simply specifies the data and markup of a page, and delegates incremental modification of pages to FORWARD's optimizations.

The page specification syntax resembles that of template engines bundled with mainstream web frameworks (such as eRuby of Ruby-on-Rails) by enclosing {\em processing directives} between {\tt <\% ... \%>} and embedding them into a HTML document. Directives supported include:

\begin{compact_enum}
\item {\bf HTML and Visual Units: } HTML is specified as-is. In addition, FORWARD provides {\em visual units wrappers} around JavaScript components of popular libraries, so that they can be specified as easily as HTML: as basic template markup, instead of via ad-hoc JavaScript code for custom integration. Visual units have state, i.e. they can store user input. (Section~\ref{sec:page-static})
\item {\bf Queries, Iteration and Conditionals: } SQL++ queries executing over the UAS are used to generate dynamic data. The {\tt for} directive instantiates markup by iterating through tables, whereas {\tt if/then/else} directives instantiate markup that are conditionally displayed. (Section~\ref{sec:page-report})
\item {\bf Events and Data Binding: } The {\tt event} directive specifies that an event fired by a HTML element or visual unit is to invoke an action. The {\tt bind} directive binds user input into the UAS, so that an action can read and write them. (Section~\ref{sec:page-event})
\end{compact_enum}

Each of the following sub-sections illustrates these directives by building an increasingly richer page towards implementing the page of Figure~\ref{fig:screenshot}.

\subsubsection{Static Page with HTML and Visual Units}
\label{sec:page-static}

\begin{figure}[htbp]
\begin{center}
\includegraphics[width=3.0in]{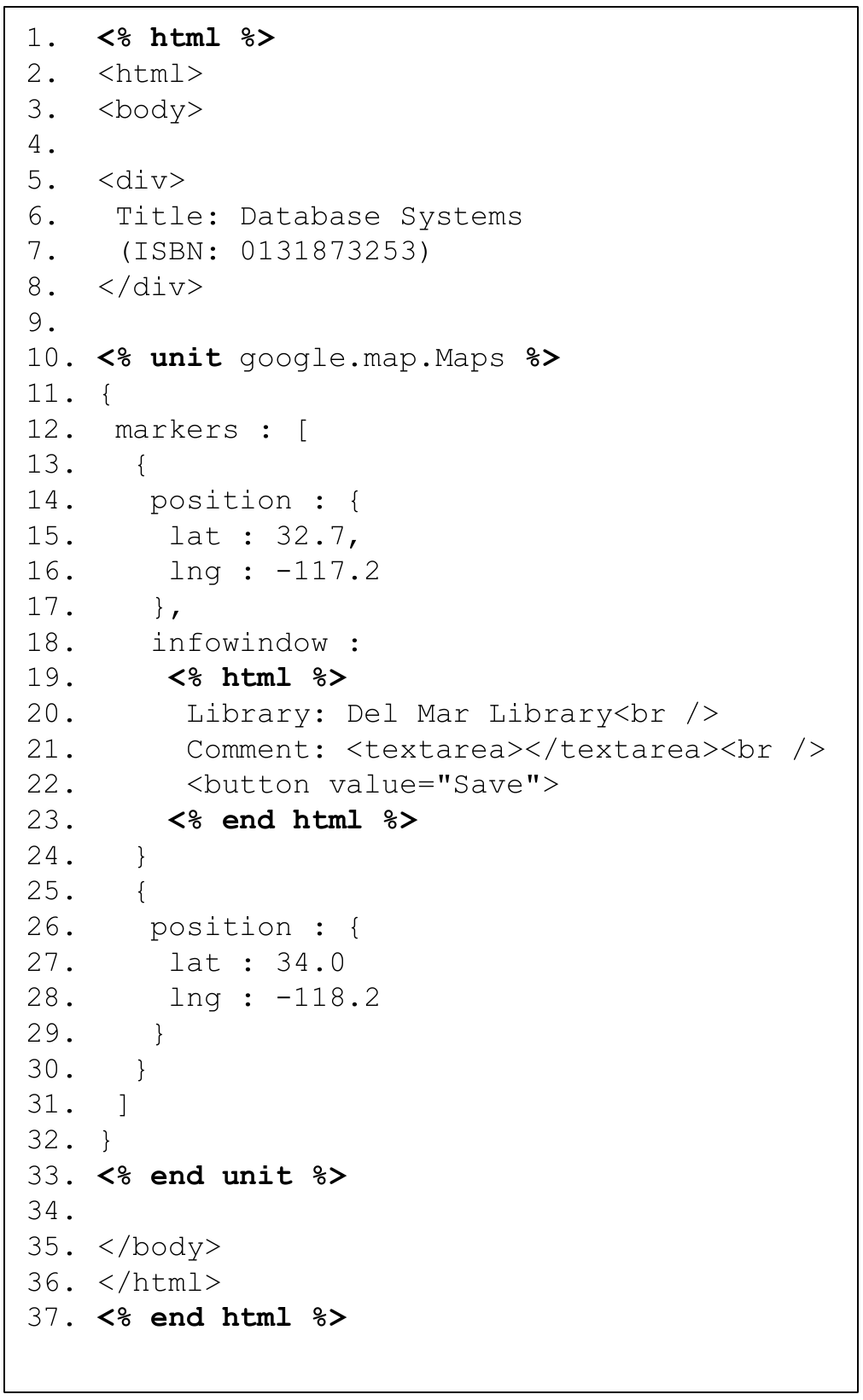}
\caption{Page specification with HTML and Visual Units}
\label{fig:page-static}
\end{center}
\end{figure}

Figure~\ref{fig:page-static} implements a page that displays HTML and a Google Maps visual unit, with all data statically specified as constants. Notice that the page contains only directives, HTML markup and JSON markup: No JavaScript code is necessary. The {\tt html} directive (Lines~1-37, 19-23) accepts any HTML markup within its body. The {\tt unit} directive (Lines~10-33) takes as argument the package and class name of a FORWARD visual unit, and accepts any valid JSON markup conforming to the schema specified by the visual unit wrapper. For example, the Google Maps visual unit has state represented by an object literal (Lines~11-32), which includes a {\tt markers} array (Lines~12-31). Even though the Google Maps unit has numerous properties that maintain state (including map type, zoom level, rotation, overlays etc.), most of them are optional and may be omitted for convenience. Both HTML and visual units can in turn contain nested HTML and visual units. For example, a marker has an optional {\tt infowindow} which contains a child {\tt html} directive (Lines~18-23). (When a marker is selected, the Google Maps component has native behavior that shows a pop-up window displaying the specified HTML.) As syntactic sugar, the top-level {\tt html} directive can be omitted, i.e. any HTML file is a valid page specification.

To support easy integration of JavaScript components through template markup, FORWARD provides pre-built visual units wrapping popular JavaScript libraries, which currently include: (a) bar charts, line charts, pie charts etc. from HighCharts and Google Visualization (b) street maps from Google Maps (c) dialogs, date pickers, pop-up menus etc. from Google Closure and (d) the Ace Editor with syntax highlighting, indentation and auto-completion.

\paragraph{Translating markup to Ajax displays} Since all data has been specified as constants in Figure~\ref{fig:page-static}, evaluating the page specification will produce an identical page instantiation. A page instantiation comprises page state and unit annotations. The page state models the data of the page instantiation in SQL++ as follows. (1) For a {\tt html} directive, its data is represented as a tuple with two values: a {\tt template} string for the HTML markup extended with placeholders for children directives, and a {\tt children} tuple for the data of children directives. (2) For a {\tt unit} directive, its data is represented with the SQL++ equivalent of the JSON markup. For example, the page state of Figure~\ref{fig:page-static} is a tuple with a {\tt template} string {\tt <html> ...<placeholder id="\_1" /> ... </html>} and a {\tt children} tuple that has a {\tt \_1} tuple corresponding to the map's data. The unit annotations of the page instantiation indicate the mapping between SQL++ values and visual unit wrappers. For example, the root tuple of the page state has unit annotation {\tt html}, whereas the {\tt \_1} tuple has unit annotation {\tt google.map.Maps}.

Whereas HTML within a browser is modeled as DOM state, most JavaScript components provide only programmatic APIs, i.e. method calls that incrementally re-render the components. The unit annotations induce a split of the page state into one sub-tree per visual unit, such that each visual unit wrapper is responsible for bidirectional synchronization between the page state of the visual unit and the corresponding DOM/JavaScript component state: (a) {\em collectors} registered as event handlers of HTML or JavaScript components are notified whenever user input occurs, and thereby propagate such changes into the page state of the visual unit (b) {\em renderers} translate diffs on the page state of the visual unit into updates of the underlying DOM elements or method calls of the underlying JavaScript components. As a result of the page state abstraction, the page specification allows JavaScript components to be specified easily using template markup, whereas the custom logic to translate page state into efficient incremental methods is delegated to visual unit wrappers.

\subsubsection{Dynamic Report with Queries and Control Flow}
\label{sec:page-report}

\begin{figure}[htbp]
\begin{center}
\includegraphics[width=3.3in]{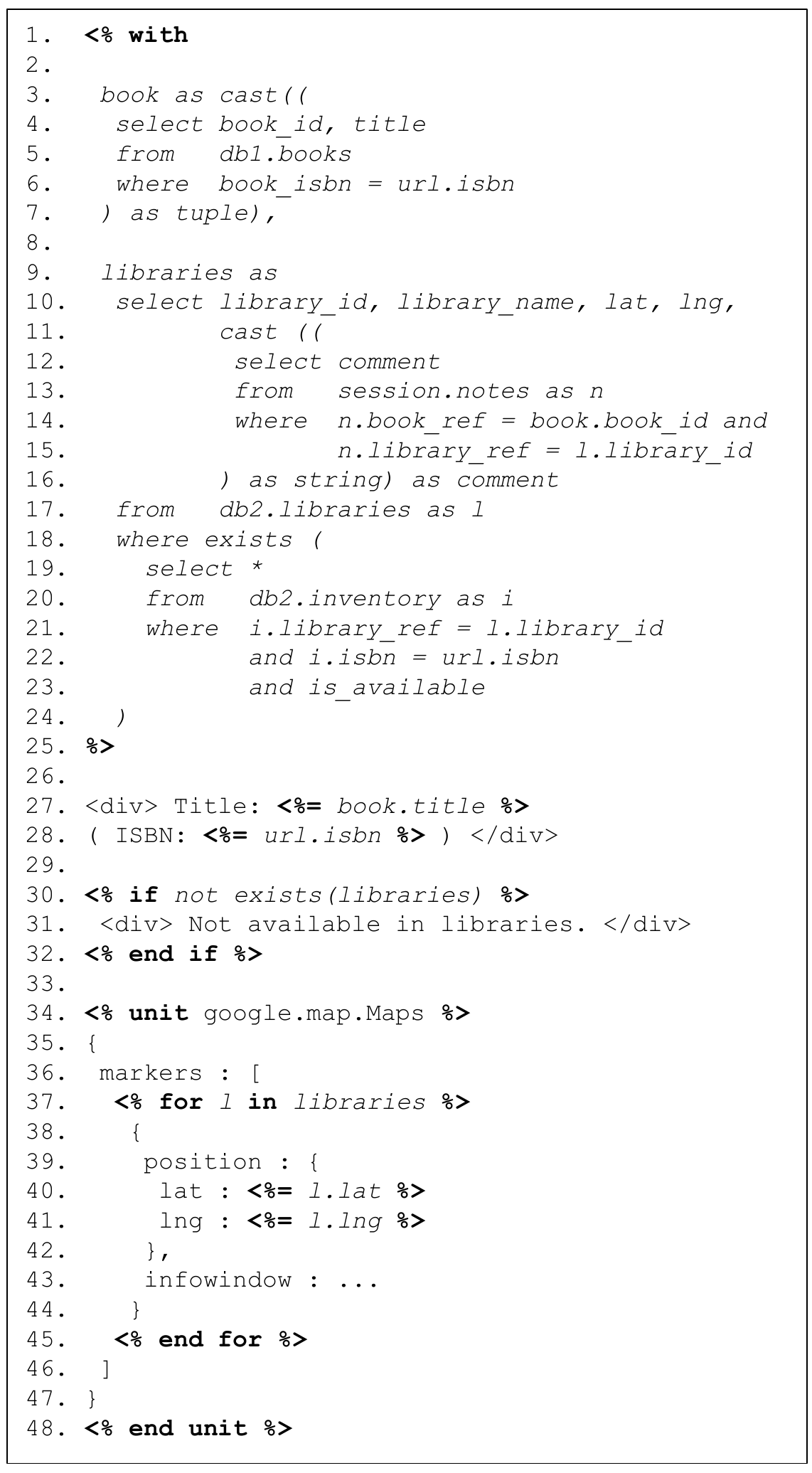}
\caption{Page specification with queries, iteration and conditionals}
\label{fig:page-report}
\end{center}
\end{figure}

Whereas the earlier section specifies a static page, typically data-driven web applications display pages with dynamic data. Figure~\ref{fig:page-report} implements a page where all data displayed are dynamically generated with SQL++ queries (highlighted in italics). In addition to reading the UAS, queries can also read local views that are lexically-scoped within the page. A view is simply a (name, query) pair.

For the sample application, the UAS has three data sources: (1) {\tt db1}, which has SQL table {\tt books(book\_id, title, isbn)}. (2) {\tt db2}, which has SQL tables {\tt libraries(library\_id, library\_name, lat, lng)} and {\tt inventory(library\_ref, isbn, copy\_id, is\_available)} (3) {\tt session}, which has SQL++ table {\tt notes(book\_ref, library\_ref, comment)}.

A {\tt with} directive is analogous to SQL's {\tt WITH} keyword: it specifies local views corresponding to intermediate data to be displayed. For example, the {\tt with} directive on Line~1 specifies a {\tt book} view, which is a tuple retrieved from the database {\tt db1} that has the same ISBN number as that of the {\tt isbn} URL parameter (Lines~3-7). Also, a {\tt libraries} view, which is a table retrieved from a second database {\tt db2} corresponding to all libraries that have at least one copy of the book available for loan (Lines~9-24). The {\tt libraries} view also retrieves the user's comments from the session (Lines~11-16). The views are lexically scoped to the closest {\tt for} directive, or the entire page specification if there is none. For example, both {\tt book} and {\tt libraries} are lexically scoped to the entire page specification.

The {\tt with} directive is useful for modularizing the page specification into two separate parts: computing the data versus instantiating the template markup. Nevertheless, queries are first-class citizens within the page specification, i.e. they can be used anywhere an expression is acceptable. For example, the query for {\tt book} (Lines~3-7) can be inlined where {\tt book} is referenced (Line~27).

The {\tt =} directive instantiates the result of the query as a value within HTML or JSON markup. For example, the HTML {\tt <div>} elements display the book's title and ISBN (Lines~27-28), whereas the map displays markers corresponding to the latitude and longitude of the corresponding libraries (Lines~40-41).

Control flow is straightforward. The {\tt if/then/else} directives evaluate the condition queries, and instantiate the corresponding branch bodies. For example, the markup in Line~31 is instantiated if the {\tt libraries} table is empty. Likewise, the {\tt for} directive iterates through the results of the query: in each iteration, it binds the tuple to the local view (i.e. iterator variable), and instantiates the body. The local view is lexically scoped to the body. For example, Lines~37-45 instantiate a marker object literal for each corresponding library {\tt l}.

By using queries and markup, the page is specified to always display the ISBN and title of the book, together with a map of all libraries with available copies. In particular, the developer does not implement any incremental modification of pages to handle what happens when any of the data dependencies are updated, e.g. when the URL parameter changes, or when the availability of a book within a library changes. Page evaluation semantics guarantee that visualizations in the page state are always refreshed when a page is evaluated.

\paragraph{Automatic optimizations enabled by declarativeness} 
A SQL++ query has the advantage of being a read-only computation with no side-effects on the UAS. Furthermore, the page specification is declarative. Both of these properties enable FORWARD to automatically perform and optimize many low-level tasks on behalf of the developer:

\begin{compact_enum}

\item {\bf Distributed query processing:} The developer writes SQL++ queries against the UAS in a {\em location transparent} fashion. For example, the queries of the {\tt with} directive input multiple data sources {\tt db1} (Line~5), {\tt db2} (Lines~17 \& 20), {\tt url} (Lines~6 \& 22) and {\tt session} (Line~13), but are nevertheless written as if they execute within a single location. The responsibility of distributed data access is delegated to FORWARD's distributed query processor, which leverages and enhances existing techniques for distributed query processing. As an example of a {\em partitioning optimization} that partitions a query into efficient underlying sub-queries, the size of intermediate results sent between the server and database tiers are minimized through successive query rewritings \cite{forward-data-access-ucsd-13}.

The distributed query processor currently does not consider partitioning optimizations that push sub-queries into the browser. This is a conservative design choice based on security concerns. For example, a query that checks whether a user has admin rights cannot be pushed as-is to the browser, otherwise a malicious user may hijack the query to bypass access control. Nonetheless, pushing safe sub-queries to the browser is active work in progress. We also plan to support {\em partitioning constraints} (similar in principle to those in Links \cite{links-fmco-06}), so that a developer can specify partitioning optimizations to only execute an annotated sub-query at a specified tier (Section~\ref{sec:future}).

\item {\bf Incremental view maintenance:} If the pages displayed in the browser before and after an action invocation are the result of the same page specification, FORWARD will utilize the prior page state to efficiently re-evaluate the SQL++ query using incremental view maintenance \cite{forward-sigmod-10}. As an example of incremental view maintenance in the relational model, given a view instance $V = R \bowtie S$, insertions on relation $R$ are modelled as a table $\triangle^{+}R$, and incremental view maintenance efficiently determines inserts on the view instance with $\triangle^{+}V = \triangle^{+}R \bowtie S$.

\item {\bf Incremental rendering:} The diffs produced by incremental maintenance can in turn be passed to the incremental renderers of visual unit wrappers to efficiently refresh the DOM and JavaScript components. As illustrated in \cite{forward-sigmod-10}, in addition to performance gains due to fewer DOM elements and JavaScript components being re-initialized, incremental rendering also delivers a better user experience by reducing flicker and preserving unsaved browser state such as focus and scroll positions.

\item {\bf Holistic query optimizations} FORWARD holistically optimizes the SQL++ queries of the page. For example, the {\tt with} directive does not specify whether the local view is virtual or materialized: it is left as an optimization opportunity to the query processor. Therefore, standard query optimizations such as pushing selections down can be applied across queries. As another example, suppose the {\tt for} directive (Lines~37-45) contains an additional {\tt =} or {\tt for} directive with an expensive query that is parameterized by local view {\tt l}. Even though the semantics suggest a naive nested loop evaluation where the expensive query is evaluated once for each outer {\tt l} tuple, FORWARD identifies a large class of SQL++ queries on the page as {\em set processable} and holistically rewrites them from {\em tuple-at-a-time execution plans} into more efficient {\em normalized-sets execution plans} \cite{forward-data-access-ucsd-13}.

\end{compact_enum}

\subsubsection{User Input with Events and Data Binding}
\label{sec:page-event}

\begin{figure}[htbp]
\begin{center}
\includegraphics[width=3.3in]{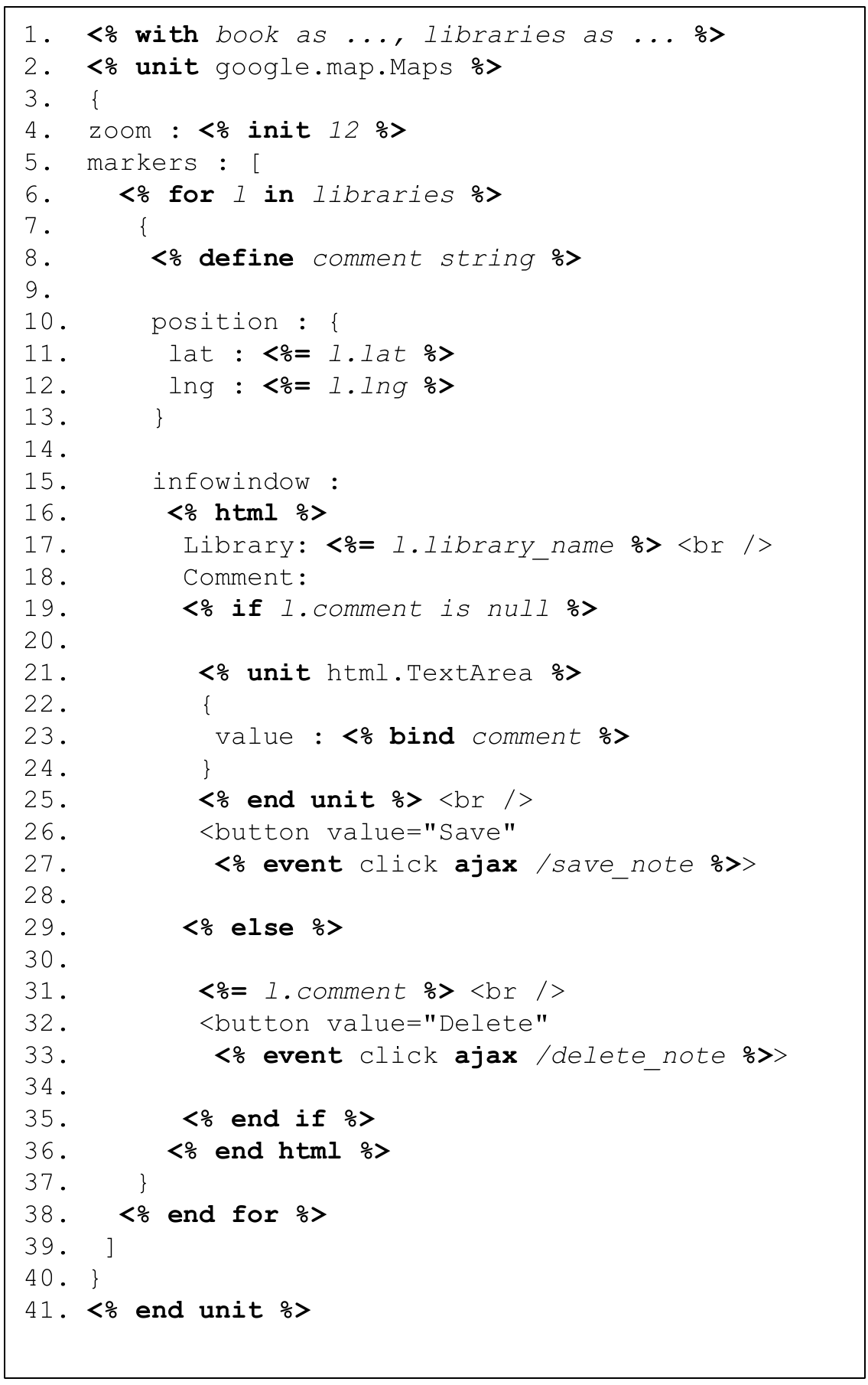}
\caption{Page specification with events and data binding}
\label{fig:page-event}
\end{center}
\end{figure}

Figure~\ref{fig:page-event} fully implements the page shown in Figure~\ref{fig:screenshot} by adding (a) initialization of the zoom level (b) the text area and {\tt Save} button (c) the text comment {\tt Delete} button. Line~19 checks whether the session does not contain a note for the particular (book, library) pair, if so the text area and {\tt Save} button are shown, otherwise the text comment and {\tt Delete} button are shown.

The {\tt init} directive specifies initialization of a value. For example, Line~4 initializes the zoom level of the map. Note the difference between using the {\tt init} and {\tt =} directives: whereas the {\tt init} directive will maintain subsequent user input that changes the zoom level, the {\tt =} directive will always overwrite user input during page evaluation.

The {\tt define} and {\tt bind} directives are closely related: together they specify data bindings, so that user input in the page state are accessible by an action. A {\tt define} directive specifies a local schema, whereas a {\tt bind} directive maps a value in the page state into the defined schema. The defined schema is lexically scoped to the closest for directive, or the entire page specification if there is none. For example, Line~8 defines a string {\tt comment} for each library, whereas Line~23 binds the text area {\tt value} of each library marker into the {\tt comment} schema.  As syntactic sugar, the {\tt define} directive can be inferred from the corresponding {\tt bind} directive, and may hence be omitted for convenience.

Both {\tt init} and {\tt bind} serve to indicate that a value corresponds to user input. The semantics of page evaluation guarantee that, unlike visualizations in the page state which are always refreshed, user input in the page state are never refreshed. It is also possible to combine both {\tt bind} and {\tt init} directives. For example, using instead {\tt bind comment init `EDIT ME' } on Line~23 will initialize the local schema {\tt comment} accordingly.

The {\tt event} directive specifies that in response to an event fired by a HTML element or visual unit, an action should be invoked via the specified method with the specified arguments. For example, Line~27 specifies that when the button is clicked, the action at URL {\tt /save\_note} should be invoked via {\tt ajax}, which refreshes the existing page without updating the URL in the address bar. Other methods of invocations are also supported. If significant portions of the existing page may be refreshed as a result of an action, it is recommended that the action not be executed concurrently with any other action: the corresponding method to specify is {\tt sync\_refresh}, which is a synchronous equivalent of {\tt ajax}. To set the address bar to URLs of other applications, in particular REST APIs of other web services, the developer can also use methods {\tt get}, {\tt post}, {\tt put} and {\tt delete} (i.e. the four HTTP verbs).

Passing parameters to an action occurs through a special {\tt request} source that automatically maps to each local view (by {\tt with} and {\tt for} directives) and local schema (by {\tt define} directive) in lexical scope. For example, the {\tt save\_note} action has access to {\tt request.book} ({\tt with} directive in Line~1), {\tt request.libraries} ({\tt with} directive in Line~1), {\tt request.l} ({\tt for} directive in Line~6) and {\tt request.comment} ({\tt define} directive in Line~8). A local view is read-only to ensure that an action cannot modify the results of any SQL++ view of the page, whereas a local schema is read-write to allow an action to initialize defaults or override user input. In particular, the mapping of the {\tt request} source guarantees that (a) when user input occurs, the updated visual value is readable by the action in the local schema (b) when an action writes to the local schema, the write will propagate automatically to the visual value mapped by {\tt bind}. This is also known as {\em two-way data binding} \cite{data-binding-microsoft-13} in UI frameworks. In the example, the entire {\tt request} is read-only, except for {\tt request.comment} which is read-write. The action can access user input in the text area by reading {\tt request.comment}; it can also overwrite user input with an {\tt UPDATE} statement on {\tt request.comment}. 

Lastly, when an action is invoked, the local views in the {\tt request} source correspond to the page state of what the user saw when user input occurred. For example, suppose the book was available at library $l_1$ when the page was displayed to the user, and while the user was typing, the book became unavailable in the database. When the user invokes the {\tt save\_note} action, $l_1$ will be included within local view {\tt request.libraries}. This design enables the action to validate user input against the data that was previously displayed.

\paragraph{Classifying page state with respect to user input} Whereas Section~\ref{sec:page-report} presents visualizations, this section presents user input. In general, a page has two parts with distinct functionalities: a form part that maintains user input (either through HTML forms or JavaScript components), and a visualization part that is displayed to the user but cannot be changed directly via user interaction. These two parts are structurally commingled: the sample application shows the {\tt markers} table which has mostly visualization parts, except for the {\tt comment} form part. It is also possible to have the opposite: consider an {\tt invoices} table with mostly form parts, except for a visualization part {\tt total price} which is a calculation over form parts {\tt unit price} $\times$ {\tt quantity}. Due to the commingling, the page state is modeled at a fine granularity such that each value is either (a) a {\em base value}, which is specified by {\tt bind} and {\tt init} directives, and is readable and writable by the user and actions or (b) a {\em derived value}, which is specified by an {\tt =} directive, and is read-only for the user and actions. Effectively, base values represent the mutable state within the page state which may be concurrently written by multiple asynchronous actions, whereas derived values represent computations within the page state that are always consistent with respect to this mutable state and the rest of the UAS. Since base values are always distinct from derived values, FORWARD guarantees that asynchronous actions will not cause visualization refreshes that conflict with user input.

\subsection{Action Specifications}
\label{sec:action}

\begin{figure}[htbp]
\begin{center}
\includegraphics[width=3.0in]{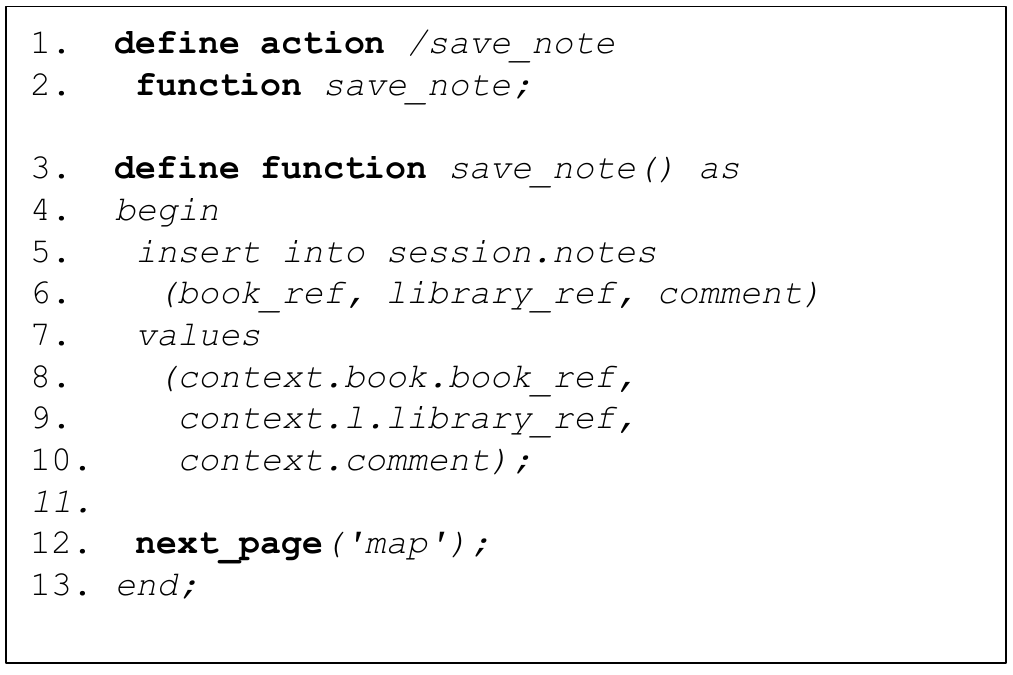}
\caption{Action specification for saving a note}
\label{fig:action}
\end{center}
\end{figure}

An {\em action specification} specifies transitions in the UAS using simple {\tt insert}, {\tt update} and {\tt delete} statements, and basic control flow statements such as conditionals, loops and functions. The syntax and semantics are closely based on PL/SQL, which is the stored procedure language of SQL databases.

Figure~\ref{fig:action} shows the action specification implementing the {\tt /save\_note} action in the sample application. An action is a mapping from a URL pattern to a function. Given an incoming URL, the runtime finds an action which has a URL pattern matching the URL, then executes the function. For example, Lines~1-2 maps the literal URL {\tt /save\_note} to the function {\tt save\_note}, which is defined in Lines~3-13. This is useful for re-using functions for different actions. As syntactic sugar, a developer can define an action mapping to an anonymous function with {\tt define action /save\_note as begin ... end}.

Within a function, {\tt insert}, {\tt delete} and {\tt update} statements read and write the UAS in a location transparent fashion. For example, Lines~5-10 show an {\tt insert} statement that reads values from the {\tt request} source, and writes them into the {\tt notes} schema of the {\tt session} source. The next page to evaluate and display is set using the special {\tt next\_page} function, which stores the page name in the UAS. For example, Line~12 specifies the next page to be the same {\tt map} page. As syntactic sugar, the {\tt next\_page} function can be omitted if the page remains the same.

Other statements supported in function specifications include: (a) {\tt declare}, which declares local SQL++ schemas (b) {\tt :=}, which assigns to schemas the results of evaluating queries (c) {\tt if/then/else} (d) {\tt for} (e) {\tt raise}, which raises exceptions and (f) {\tt exception}, which performs exception handling.

Function arguments are passed by value as in PL/SQL. To indicate if a function has side-effects, a developer specifies whether a function is {\tt immutable} or {\tt mutable}: only {\tt immutable} functions can be used in the queries of pages. The query compiler enforces that {\tt immutable} functions cannot contain {\tt insert}, {\tt delete} and {\tt update} statements.

\paragraph{Execution semantics of actions} The runtime provides the following guarantees when executing actions: (1) A synchronous action (i.e. specified with {\tt sync\_refresh}) will not be executed concurrently with other actions. (2) An action is executed only after the page state has been fully synchronized with the browser state up to the point of the event being fired, so that the action accesses only an up-to-date UAS. (3) While an action executes, it reads and writes an isolated snapshot of the page state, so that it is unaffected by other concurrent action executions. (4) All writes that an action performs on the page state are deferred until after page evaluation, upon which the runtime replays the writes in a single batch before page evaluation. (5) If an action's writes conflict with those of other actions and/or user input, the runtime non-deterministically chooses the winning writes%
\footnote{This is the behavior exhibited by conventional Ajax applications, but a desirable improvement for future work is for the developer to specify conflict resolution policies that are specific to the application/pages (Section~\ref{sec:future}).}.

\section{Related Work}
\label{sec:related}

\paragraph{Abstractions over three tiers}
Using a single language as a holistic abstraction over the three tiers of web applications is the core premise of Links \cite{links-fmco-06} from the programming language community, and Hilda \cite{hilda-icde-06} from the database community. More recently in industry, Meteor \cite{meteor-13} has also adopted this approach using JavaScript as the single language. As in FORWARD, impedance mismatch is resolved by compiling or interpreting the single language into the underlying language of each tier: JavaScript in the browser, a programming language of choice in the application server, and the query language of the database (typically SQL).

Links is a strict statically-typed functional language that supports both location transparency and incremental modification of pages. (1) For location transparency between the browser and server, a Links function is optionally annotated with a partitioning constraint of either {\tt server} or {\tt client} (browser). The system enforces these constraints, and implements this abstraction by passing continuations (i.e. intermediate execution state of functions) between server and browser. For location transparency between the server and database, Links supports a limited subset of SQL (e.g. no {\tt group by}), but it is not clear if the compiler performs partitioning optimizations across in-memory variables and database tables. In contrast, FORWARD observes that most data are persisted in the database, and thus focuses on partitioning optimizations between the server and database for the full scope of SQL in order to minimize the size of intermediate results fetched from the database. Nonetheless, we plan to adopt Links' partitioning constraints and browser-side computation in future work (Section~\ref{sec:future}). (2) Links does not distinguish between actions versus pages: updating the browser occurs through functions with side-effects on the DOM, and automatic incremental computations are supported through functional reactive programming (FRP). We conjecture that the incremental view maintenance techniques utilized in FORWARD for automatic incremental modification of pages is a specialization of FRP. (3) FORWARD also inter-operates with popular third-party libraries of JavaScript components and their JSON values, whereas Links focuses on language facilities for building re-usable components from scratch using basic HTML elements.

Hilda is a declarative SQL-based language where an application is modelled with a tree of application units (AUnits). An AUnit combines simultaneously UAS, action and page functionalities: the hierarchy of AUnit schemas corresponds to the UAS, the hierarchy of AUnit SQL queries determine the data shown on the page, and the hierarchy of handlers determine the action performed. Similar to FORWARD, Hilda provides location transparency across three tiers. Furthermore, Hilda performs partitioning optimizations across the browser and server by using a cost model to find an optimal cut of the AUnit tree. However, it is not clear whether the optimal cut may push security-sensitive computation into the browser. Since Hilda uses SQL and a relational data model, it is amenable to several FORWARD techniques, including incremental view maintenance, incremental rendering, distributed query processing, page state representation of JavaScript components, splitting page state into base versus derived values etc.

Meteor is an open source web framework that uses JavaScript as the single language and JSON as the unifying data model. This is enabled by Node.js, which is an application server that executes JavaScript, and MongoDB, which is a database for JSON values. Meteor does not offer location transparency, but provides an alternative programming model based on data synchronization. The developer specifies the data that a browser-side cache subscribes to, and Meteor automatically synchronizes this cache. Cache reads have zero network latency, whereas cache writes are asynchronously sent to the server, which enforces security policies before relaying writes to the database. Ultimately, the developer is still responsible for distributed data access, because he has to partition code according to browser, server and database. For automatic incremental modification of pages, Meteor performs change propagation using data dependencies, but it is unclear whether the scope of these change propagations is close to incremental view maintenance or FRP. Also, Meteor's template language supports only HTML, and JavaScript components can only be instantiated through programmatic APIs.

\paragraph{Abstractions over two tiers} 
Web frameworks such as Echo2 \cite{echo-09} and ZK \cite{zk-09} offer Java as a single language that abstracts over the browser and server tiers. The developer writes Java to update a page in a location transparent fashion. During runtime, Java is executed on the server to update a server-side mirror of the page state, thereafter the page state is synchronized to the client. FORWARD adopts a similar mirroring approach in its implementation: the server-side page state is accessed (through the {\tt request} source) by SQL++ queries, whereas the client-side page state is accessed by renderers and collectors. However, Echo2 and ZK do not enforce pages to be side-effect free, thus developers have to manually implement incremental modification of pages.

Google Web Toolkit (GWT) \cite{gwt-09} is a web framework that similarly offers Java as a single language over the browser and server tiers, primarily enabled by its Java-to-JavaScript cross-compiler. GWT offers more flexibility than Echo2 and ZK because computation can be selectively pushed into the browser. Unfortunately, the developer has to partition code between the browser and server, and perform distributed data access through GWT's RPC mechanism. Similar to Echo2 and ZK, a developer also manually implements incremental modification of pages.

Object-Relational Mapper (ORM) libraries such as ActiveRecord \cite{active-record-13}, Hibernate \cite{hibernate-13} and Entity Framework \cite{entity-framework-sigmod-07} provide abstractions between the server and database tiers. Arguably, they have mixed success in providing a single language abstraction due to the well-known and prevailing impedance mismatch between databases and programming languages \cite{impedance-mismatch-odbms-05}. For the subset of SQL where the object navigation abstraction works well (i.e. foreign-primary key joins, selection on primary keys), ORMs indeed offer location transparency for the developer. Recent extensions to mainstream ORMs \cite{avalanche-safe-pvldb-10,switch-icde-12} have gone further to provide partitioning optimizations for data access spanning across in-memory objects and database tables.

\paragraph{Abstractions within a single tier}

Flapjax \cite{flapjax-sigplan-2009} is a language that compiles into JavaScript, and provides automatic incremental modification of pages through functional reactive programming (FRP). The language offers primitives for event streams and behaviors, which allows the developer to specify pages that are reactive. Since values are automatically updated when their data dependencies change, developers do not need to provide code for incremental modification of pages. Flapjax's reactive semantics also apply when integrating JavaScript components from third-party libraries. However, since Flapjax is browser-centric, it is orthogonal to incremental computation in the server and database tiers.

Client-side web frameworks such as AngularJS \cite{angular-13} and KnockoutJS \cite{knockout-13} provide two-way data binding. Unlike traditional MVC frameworks where only the UI view has a data dependency on the model, two-way data binding enables the model to also have a data dependency on the UI view. This is particularly useful for UI components that accept user input. For example, given a text box with a two-day data binding to a string, user input in the text box will propagate to the string, and programmatic updates to the string will be displayed in the text box. FORWARD similarly adopts two-way data binding for base values in the page state, thus user input is accessible by actions and the writes of actions are displayed to users.

\section{Experience and Future Work}
\label{sec:future}

Based on our experience with implementing commercial and academic applications in FORWARD, we have discovered assumptions that were too stringent and require future work.

\paragraph{Dynamic queries} Pages are currently specified with statically known queries. In applications where queries need to be generated dynamically, such as a search interface that allows users to create custom filter conditions, this functionality need to be simulated through Java UDFs. A baseline solution is to dynamically create SQL++ queries by constructing strings. Higher-level APIs can also be provided by observing that typically, while the dynamism appears in selection conditions, the projection list is statically known.

\paragraph{Lightweight integration of JavaScript components} Implementing the renderers and collectors of visual unit wrappers requires big upfront time investment, particularly for components with huge APIs comprising 50 or more methods. Moreover, libraries such as D3 that provide APIs for creating custom visualizations are immensely popular, but their inherent flexibility makes it hard to create corresponding visual unit wrappers. Both problems can be simultaneously addressed by providing a lightweight API that enables integrating JavaScript components directly through the page specification without visual unit wrappers. A developer can specify JSON markup in the page specification, and map respective values to custom incremental rendering functions that he implements.

\paragraph{Maximum automatic action parallelism}
Although asynchronous actions provide ample opportunities for low latency through parallelism, there is significant risk of invalid actions being issued if their executions are interleaved while not truly independent of each other. A possible solution is to perform an automatic static analysis of the dependencies between actions and pages due to the declarative nature of SQL nature. Thus, the compiler of the system can produce warnings for dependencies detected between actions and fragments of pages, and provides suggestions to the developer on how to synchronize actions where one depends on the other.

\paragraph{Low latency and browser-side computation} While location transparency is currently achieved by mirroring page data onto the server, alternate methods are also possible. Pushing computation to the browser-side will enable low latency, particularly if an action/page requires only page data and does not need to incur a browser-server roundtrip. Another possibility is to create caches of parts of the application state on the browser, so that page evaluation can display early results from the cache, and refine the display after communicating with the server and database. FORWARD should perform automatic partitioning optimizations of the actions and pages, while allowing the developer to specify partitioning constraints to further restrict how actions and pages are partitioned.

\bibliographystyle{abbrvnat}
\bibliography{main}

\clearpage

\appendix
\section{Full specifications of running example}

\begin{figure}[h!]
\centering
\includegraphics[width=3.3in]{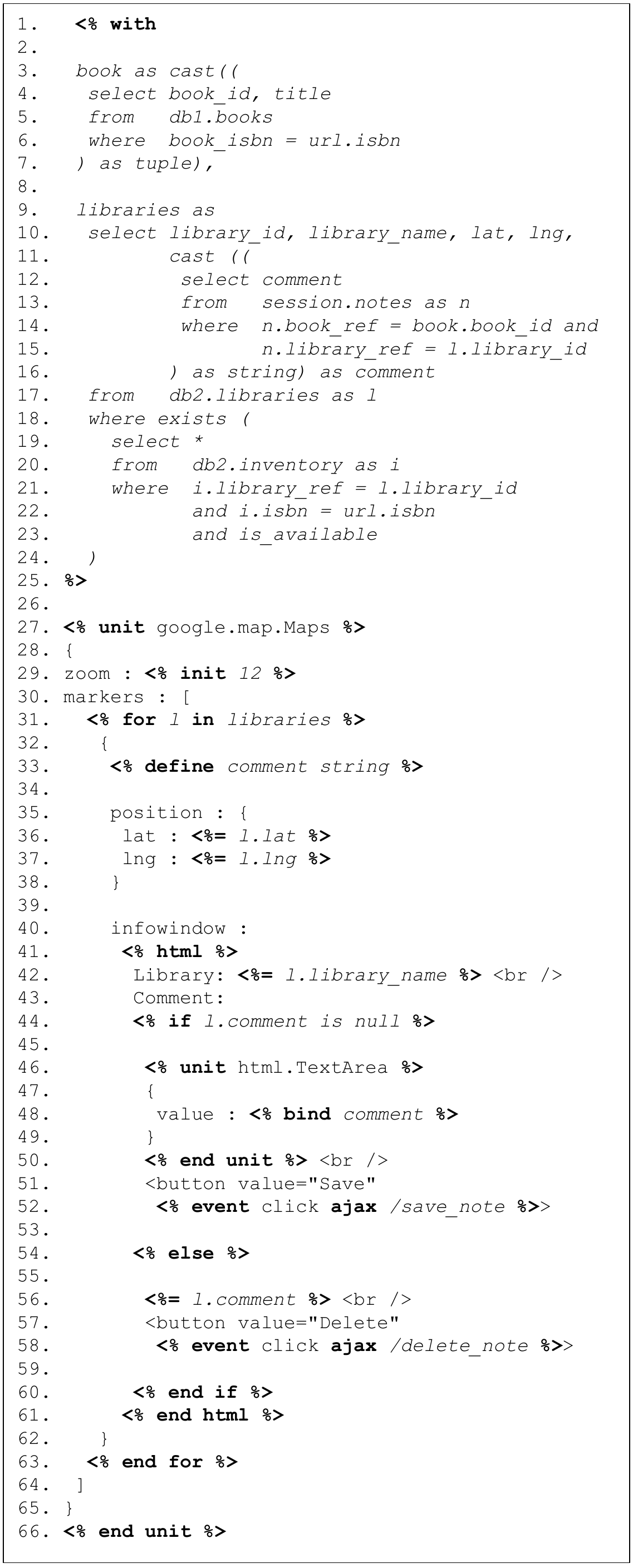}
\caption{Page specification for \texttt{map}}
\label{fig:appendix-map-page}
\end{figure}

\begin{figure}[h!]
\centering
\includegraphics[width=3.0in]{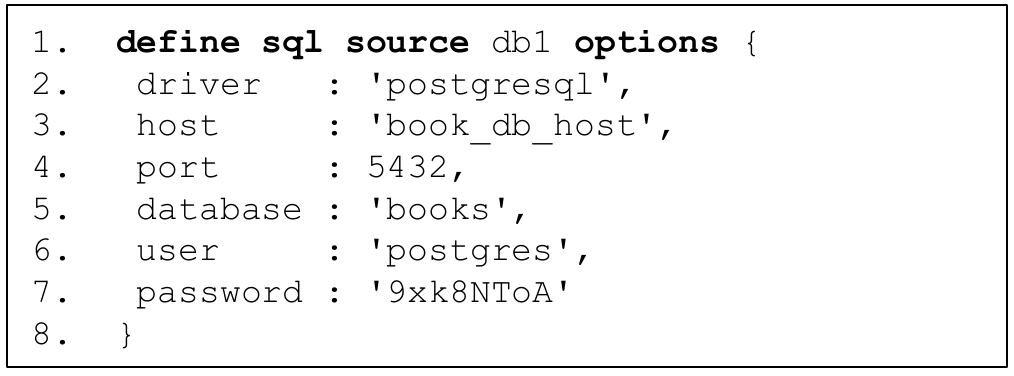}
\caption{Source specification for \texttt{db1}}
\label{fig:appendix-db1-source}
\end{figure}

\begin{figure}[h!]
\centering
\includegraphics[width=3.0in]{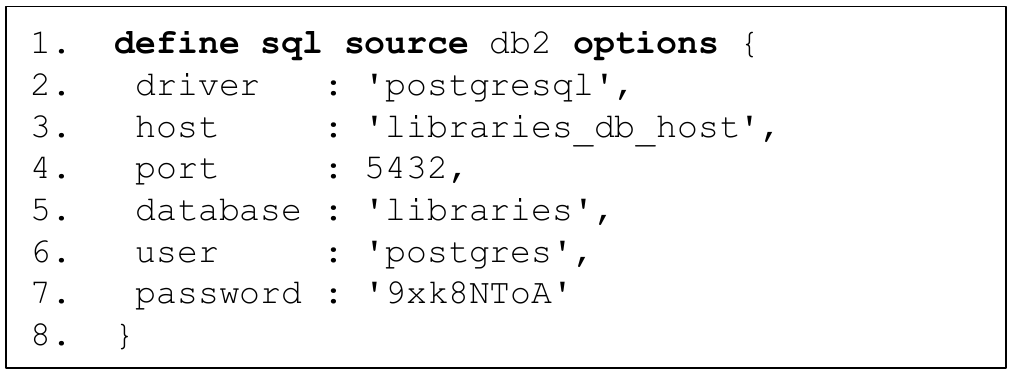}
\caption{Source specification for \texttt{db2}}
\label{fig:appendix-db2-source}
\end{figure}

\begin{figure}[h!]
\centering
\includegraphics[width=3.0in]{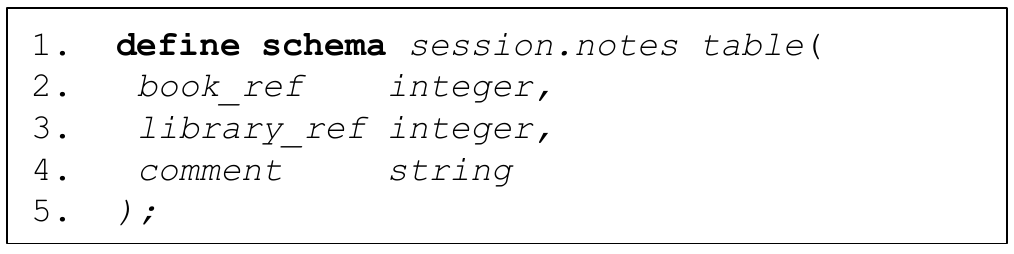}
\caption{Schema specification for \texttt{session}}
\label{fig:appendix-session-schema}
\end{figure}

\begin{figure}[h!]
\centering
\includegraphics[width=3.0in]{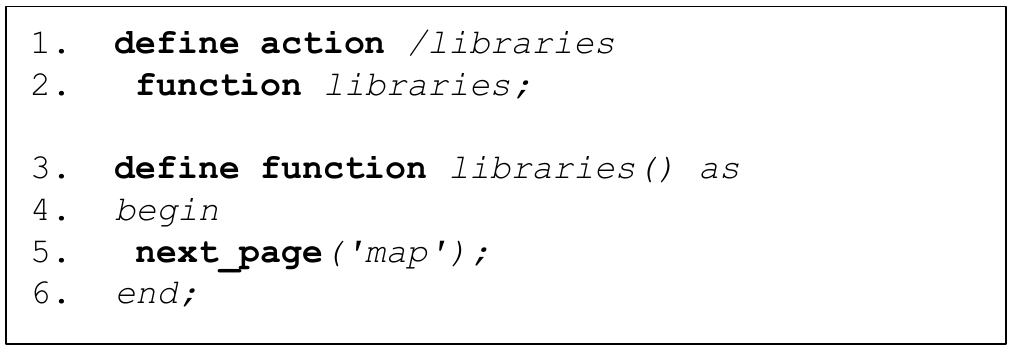}
\caption{Action specification for \texttt{/libraries}}
\label{fig:appendix-libraries-action}
\end{figure}

\begin{figure}[h!]
\centering
\includegraphics[width=3.0in]{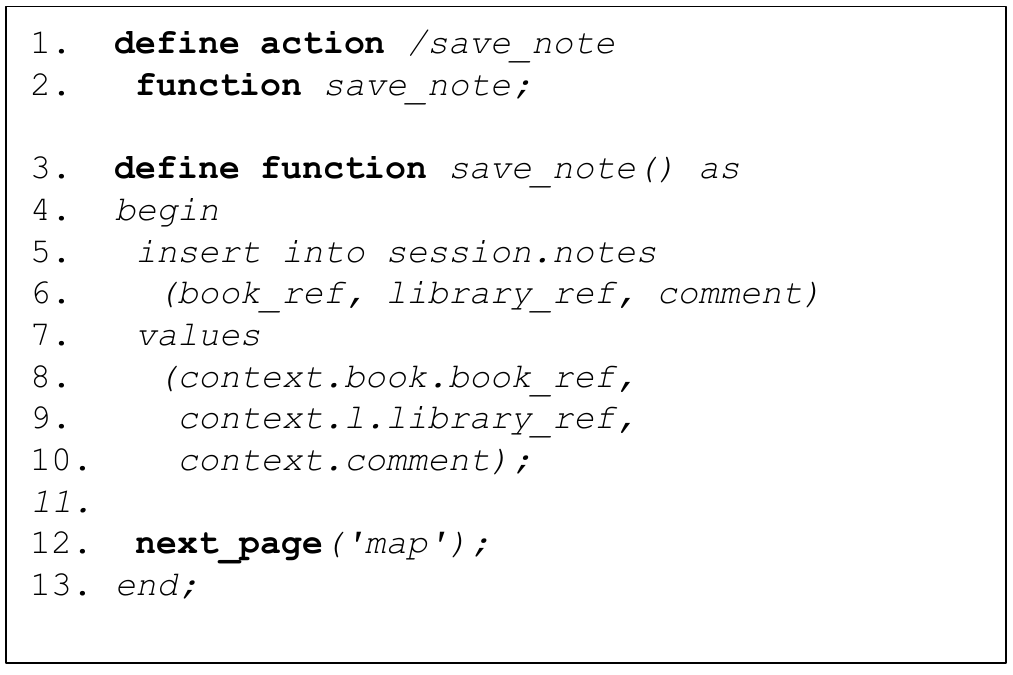}
\caption{Action specification for \texttt{/save-note}}
\label{fig:appendix-save-note-action}
\end{figure}

\begin{figure}[h!]
\centering
\includegraphics[width=3.0in]{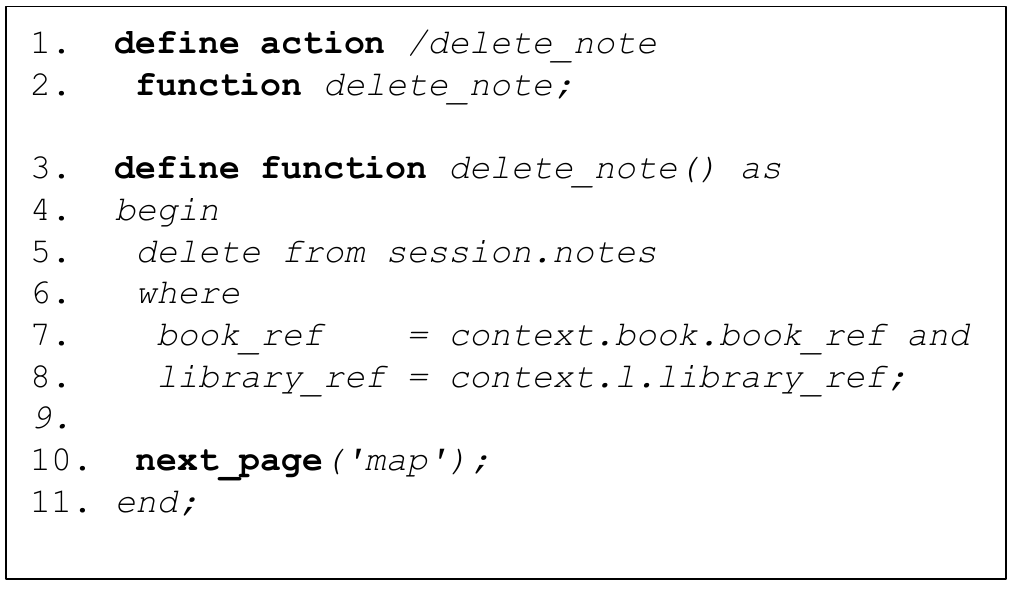}
\caption{Action specification for \texttt{/delete-note}}
\label{fig:appendix-delete-note-action}
\end{figure}

\end{document}